\documentclass[twocolumn,showpacs, 
amsmath,amssymb, showkeys]{revtex4}

\usepackage{graphicx}
\usepackage{dcolumn}
\usepackage{bm}

\begin{document}


\title{Evolution of Quantum Fluctuations Near the Quantum Critical Point\\
of the Transverse Field Ising Chain System CoNb$_2$O$_6$}

\author{A. W. Kinross$^{1}$, M. Fu$^{1}$, T. J. Munsie$^{1}$, H.\ A.\ Dabkowska$^{2}$, G. M. Luke$^{1,3}$, Subir Sachdev$^{4}$, and T. Imai$^{1,3}$} \email{imai@mcmaster.ca}

\affiliation{$^{1}$Department of Physics and Astronomy, McMaster University, Hamilton L8S4M1, Canada} 
\affiliation{$^{2}$Brockhouse Institute for Materials Research, McMaster University, Hamilton L8S4M1, Canada} 
\affiliation{$^{3}$Canadian Institute for Advanced Research, Toronto, Ontario M5G1Z8, Canada}
\affiliation{$^{4}$Department of Physics, Harvard University, Cambridge, Massachusetts 02138, USA}

\date{\today}

\begin{abstract}
The transverse field Ising chain (TFIC) model is ideally suited for testing the fundamental ideas of quantum phase transitions, because its well-known $T=0$ ground state can be extrapolated to finite temperatures.  Nonetheless, the lack of appropriate model materials hindered the past effort to test the theoretical predictions.  Here we map the evolution of quantum fluctuations in the TFIC based on Nuclear Magnetic Resonance (NMR) measurements of CoNb$_2$O$_6$, and demonstrate the finite temperature effects on quantum criticality for the first time.  From the temperature dependence of the $^{93}$Nb longitudinal relaxation rate $1/T_1$, we identify the renormalized classical, quantum critical, and quantum disordered scaling regimes in the temperature ($T$) vs. transverse magnetic field ($h_{\perp}$) phase diagram.  Precisely at the critical field $h_{\perp}^{c}=5.25 \pm 0.15$ T, we observe a power-law behavior, $1/T_{1} \sim T^{-3/4}$, as predicted by quantum critical scaling.  Our parameter-free comparison between the data and  theory reveals that quantum  fluctuations persist up to as high as $T \sim 0.4 J$, where the intra-chain exchange interaction $J$ is the only energy scale of the problem. 
\end{abstract}

\maketitle

\section{\label{sec:level1}Introduction}
The concept of $T=0$ quantum phase transitions has emerged as an overarching theme in strongly correlated electron physics \cite{Hertz, Chakravarty, Millis, Chubukov, Goldman, Sachdev, Sachdev2, Si, Sachdev3}.  The nature of quantum fluctuations near the quantum critical point, however, remains enigmatic \cite{Lonzarich}.  How well does the quantum criticality account for finite temperature properties?  How high in temperature does the effect of the quantum critical point persist?\cite{Lonzarich, Chakravarty2}  Do quantum fluctuations remain strong enough at elevated temperatures to account for the mechanism of exotic superconductivity in copper oxides, iron pnictides, and heavy Fermions systems?  The dearth of appropriate model materials for rigorously solvable Hamiltonians has not permitted experimentalists to address these fundamental questions concretely, even for the transverse field Ising chain (TFIC) \cite{Pfeuty}, a celebrated textbook example of quantum criticality \cite{Sachdev}.  Very recently, the Ising chain material CoNb$_2$O$_6$ \cite{Maartense, Scharf, Ishikawa, Heid, Kobayashi, ESR, ESR2} was proposed to be an ideal model system of the TFIC based on neutron scattering measurements in transverse magnetic fields \cite{Coldea}, paving a new avenue to investigate the finite temperature effects on quantum fluctuations in the vicinity of a quantum critical point (QCP).

The TFIC Hamiltonian is deceptively simple \cite{Pfeuty, Sachdev};
\begin{equation}
H = -J \sum_{i}(\sigma_{i}^{z}\sigma_{i+1}^{z} + g \sigma_{i}^{x}),
\end{equation} where $J$ ($>0$ for ferromagnetic Ising chains in CoNb$_2$O$_6$) represents the nearest-neighbor spin-spin exchange interaction, $\sigma_{i}^{z(x)}$ is the {\it z}({\it x})-component of the Pauli matrix at the {\it i}-th site, and the dimensionless coupling constant $g$ is related to the transverse magnetic field $h_{\perp}$ applied along the {\it x}-axis as $g = h_{\perp}/h_{\perp}^{c}$, where $h_{\perp}^{c}$ is the critical field ($h_{\perp}^{c} = 5.25 \pm 0.15$ Tesla in CoNb$_2$O$_6$, as shown below).  Since $\sigma_{i}^{z}$ and $\sigma_{i}^{x}$ do not commute, the classical Ising Hamiltonian for $g=0$ becomes the quantum TFIC Hamiltonian for $g >0$.  The QCP is located at $g=1$, where the applied field is tuned precisely at $h_{\perp}^{c}$; a magnetic field greater than $h_{\perp}^{c}$ coerces the magnetic moments along its direction and transforms the $T=0$ ferromagnetic ground state to a paramagnetic state.  See Fig.\ 1 for the generic theoretical phase diagram of the TFIC \cite{Sachdev, Young}.  In spite of its apparent simplicity, the TFIC served as the foundational model for quantum Monte Carlo simulations \cite{Suzuki}, and continues to attract attention in quantum information theory \cite{Waterloo}.   

A major advantage of working with the TFIC as a model system for testing the fundamental ideas of quantum phase transitions is that, in the absence of a transverse magnetic field ($g=0$), the thermodynamic properties of the Ising chain can be rigorously solved at arbitrary temperatures \cite{Textbook}.  Even in a finite transverse field ($g>0$), the TFIC is well understood at $T=0$ \cite{Pfeuty, Suzuki, IndianBook}, and QC (Quantum Critical) scaling theory extended the $T=0$ results to finite temperatures \cite{Sachdev, Young}.    

We show the crystal structure of CoNb$_2$O$_6$ in Fig.\ 2 \cite{Structure}. All the pictorial images of the crystal structure in this paper were drawn using VESTA \cite{Momma}.  The Co-O-Co chains propagate along the c-axis, and the easy axis of the Co moments lies within the ac-plane \cite{Scharf,Ishikawa}.  The ferromagnetic super-exchange interaction between the nearest-neighbor Co ions is estimated to be $J =17\sim 23$ K, based on ESR \cite{ESR} and neutron scattering \cite{Coldea} measurements.  From the disappearance of magnetic Bragg peaks in the transverse magnetic field applied along the b-axis, the three-dimensional (3D) critical field was estimated to be $h_{\perp}^{c, 3D} = 5.5$ Tesla \cite{Coldea, Wheeler}.  The inter-chain couplings between adjacent Co chains are antiferromagnetic \cite{Scharf, Coldea}, weaker than $J$ by an order of magnitude \cite{ESR, Coldea}, and frustrated \cite{Scharf, Balents}.  This means that the 3D magnetic long range order induced by inter-chain interactions, which tends to mask the effects of the one dimensional (1D) QCP of the individual Ising chains, is suppressed; the 3D ordering temperature is as low as $T_{c}^{3D}=2.9$ K even in $h_{\perp} = 0$ \cite{Scharf,Ishikawa}.   Combined with the modest $J$, Ising chains in CoNb$_2$O$_6$ are ideal for testing the TFIC Hamiltonian, but were overlooked for three decades. 

In what follows, we will report $^{93}$Nb NMR (Nuclear Magnetic Resonance) investigation of quantum spin fluctuations in CoNb$_2$O$_6$.  NMR is a powerful low energy probe, and good at probing the physical properties near QCP's \cite{Imai1993, Imai2, Imai3, Grenoble, BoseEinstein, Reyes, Ning, Nakai, Zheng}.  We will map the evolution of low energy quantum fluctuations of Co spins near the QCP, by taking advantage of the hyperfine interactions between Co electron spins and $^{93}$Nb nuclear spins.  We will experimentally verify the phase diagram of the TFIC in Fig.\ 1 above $T=0$ for the first time, and demonstrate that the effect of the QCP persists at finite temperatures as high as $T \sim 0.4 J$.

\section{\label{sec:level1}Experimental}
We grew the CoNb$_{2}$O$_{6}$ single crystal from a stoichiometric mixture of cobalt and niobium oxides using a floating zone furnace.  We assessed the surface quality and oriented the crystal utilizing Laue x-ray diffractometry.  Once the material was sectioned into oriented slices along the a, b and c crystallographic directions, these were individually scanned with the Laue diffractometer and showed a uniform, single-crystalline structure. A small section of the single crystal was ground into a powder and analyzed using powder x-ray diffraction which showed only single phase cobalt niobate in the crystal within instrument resolution.  The features present in the SQUID magnetometry data shown in Fig.\ 2(d) matched previously published data on this material \cite{Scharf}.  

For NMR measurements, we cut a piece of single crystal with the approximate dimensions of 4 mm x 2 mm x 5 mm.  We glued the crystal to a sturdy sample holder made of machinable aluminum-oxide (MACOR ceramic) with a thickness of $\sim 3$ mm to ensure that the crystal orientation did not change at low temperatures.  We found that the strong magnetic torque applied to the crystal by the external magnetic field could easily bend sample holders made of soft materials such as plexiglass or plastic, and introduce noticeable systematic errors below $\sim 10$~K. 

We observed $^{93}$Nb NMR in a broad range of temperature from 2~K ($\sim 0.1 J$) up to 295~K.  We show the typical $^{93}$Nb NMR spectrum in the inset of Fig.\  3.  Since the $^{93}$Nb nuclear spin is $I=9/2$, we observed 4 pairs of satellite transitions split by a quadrupole frequency $\nu_{Q}^{b} = 1.9$ MHz, in addition to the large central peak arising from the $I_{z} = +\frac{1}{2}$ to $-\frac{1}{2}$ transition.  In the main panel of Fig.\ 3, we also show the temperature dependence of the central transition in $h_{\perp} = 5.3$ Tesla applied along the b-axis.   

We measured the $^{93}$Nb longitudinal relaxation rate $1/T_{1}$ by applying an inversion $\pi$ pulse prior to the $\pi/2 - \pi$ spin echo sequence, and monitoring the recovery of the spin echo intensity $M(t)$ as a function of the delay time $t$.  The typical width of the $\pi/2$ pulse was $\sim 1\ \mu$s.  We fit these recovery curves to the solutions of the rate equation \cite{Narath}:    
\begin{equation}
M(t) = M(\infty) - A \sum_{j=1}^{9} a_{j} e^{-b_{j} t/T_{1}},
\end{equation}
with three free parameters: $M(\infty)$, $A$, and $1/T_{1}$.  By solving the coupled rate equations for $I = \frac{9}{2}$ under the appropriate initial condition, one can calculate and fix the coefficients as $(a_{1}, a_{2}, a_{3}, a_{4}, a_{5}, a_{6}, a_{7}, a_{8}, a_{9}) = (0.653, 0, 0.215, 0, 0.092, 0, 0.034, 0, 0.06)$ for the central transition and (0.001, 0.0112, 0.0538, 0.1485, 0.2564, 0.2797, 0.1828, 0.0606, 0.0061) for the $I_{z} = \pm \frac{7}{2}$ to $I_{z} = \pm \frac{9}{2}$  fourth satellite transitions, while $(b_{1}, b_{2}, b_{3}, b_{4}, b_{5}, b_{6}, b_{7}, b_{8}, b_{9}) = (45, 36, 28,21,15, 10, 6, 3,1)$ for both cases \cite{Narath}.  

An example of the signal recovery of the central transition observed at 130\ K in $h_{\perp} = 3$ Tesla is shown in Fig.\ 4, in comparison to that observed for a fourth satellite transition on the higher frequency side.  Our results in Fig.\ 4 confirm that the best fit values of $1/T_{1}$ agree within $\sim 2$ \% between the central and satellite transitions.    The central transition is the strongest among all 9 peaks as shown in the inset of Fig.\ 3, and hence most advantageous in terms of the signal intensity.  When the relaxation rate exceeds $1/T_{1} \sim 2 \times 10^{3}$ s$^{-1}$, however, accurate measurements of $1/T_{1}$ using the central transition become increasingly difficult because the recovery curve $M(t)$ is dominated by two extremely fast normal modes, $0.653\  e^{-45t/T_{1}} + 0.215\  e^{-28t/T_{1}}$; the signal intensity, $M(t)$, begins to recover at a time scale comparable to the inversion pulse width.  Accordingly, measurements of $1/T_{1}$ using the fourth satellite transition become more advantageous in the low temperature, low field regime, because its recovery curve is dominated by slower normal modes, $0.256\  e^{-15t/T_{1}} + 0.279\  e^{-10t/T_{1}}$.   We present an additional example of the $1/T_{1}$ measurement using the fourth satellite at 2 K and $h_{\perp}=5.2$\ Tesla in Fig.\ 4.   

\section{\label{sec:level1}Results and discussions}

\subsection{\label{sec:level2}$T$ and $h_{\perp}$ dependences of $1/T_{1}$}
In Fig.\ 5, we summarize the $T$ and $h_{\perp}$ dependences of $1/T_{1}$.  Notice that $1/T_{1}$ varies by more than three orders of magnitude between $h_{\perp} = 3$ and 9 T.  Quite generally, $1/T_{1}$ probes the wave vector ${\bf k}$-integral within the first Brillouin zone of the dynamical spin structure factor $S({\bf k}, \omega_{n})$ at the NMR frequency $\omega_{n}/2 \pi$ ($\sim 50$ MHz): 
\begin{equation}
	1/T_{1} = \sum_{\bf k} |a_{hf}|^{2} S({\bf k}, \omega_{n}),
\label{2}	
\end{equation}
where $a_{hf}$ is the hyperfine coupling between the observed nuclear spin and Pauli matrices.  In essence, $1/T_{1}$ measures the strength of Co spin fluctuations at the time scale set by the NMR frequency.  

Our $1/T_{1}$ data in Fig.\ 5 exhibits two distinct field regimes at low temperatures, because the spin excitation spectrum changes its character across $h_{\perp}^{c}$, as summarized in Fig.\ 6.  Below $h_{\perp}^{c} \sim 5.3$ Tesla, $1/T_{1}$ diverges gradually toward $T=0$, signaling the critical slowing down of Co spin fluctuations in the RC (Renormalized Classical \cite{Chakravarty}) regime of Fig.\ 1 toward the $T=0$ ferromagnetic ground state of each individual Ising chain.  In other words, the spectral weight of the Co spin-spin correlation function grows at the quasi-elastic peak located at $k=0$ in Fig.\ 6(a) below $h_{\perp}^{c} \sim 5.3$ Tesla.  The Co spin-spin correlation length $\xi$ along the chain grows as $\xi \sim \exp(+\Delta/T)$ in the RC regime \cite{Sachdev}, where $\Delta$ is the gap in the spin excitation spectrum as defined in Fig.\ 6(a).  Accordingly, we expect $1/T_{1} \sim \exp(+\Delta/T)$ for $T \ll \Delta$.  We summarize the details of the theoretical expressions of $1/T_{1}$ for the TFIC in Appendix A.

In contrast, $1/T_{1}$ observed above $h_{\perp}^{c} \sim 5.3$ Tesla saturates and begins to decrease with temperature.  We recall that the $T=0$  ground state remains paramagnetic in the QD (Quantum Disordered) regime above $h_{\perp}^{c}$, as shown in Fig.\ 1, and hence there is no quasi-elastic mode of spin excitations in Fig.\ 6(b).  The latter implies that $1/T_{1}$ in the QD regime is dominated by the thermal activation of spin excitations across the gap, $|\Delta|$.  Therefore we expect $1/T_{1} \sim \exp(-|\Delta|/T)$ for $T \ll |\Delta|$.  We have thus identified the 1D QCP (one dimensional QC point)  of each individual Ising chain as $h_{\perp}^{c} \sim 5.3$ Tesla.

\subsection{\label{sec:level2}Estimation of the Spin Excitation Gap $\Delta$}
In Fig.\ 7(a), we present the exponential fit of $1/T_{1} \sim \exp(\Delta/T)$ with $\Delta$ as a free parameter.  We summarize the $h_{\perp}$ dependence of $\Delta$ in Fig.\ 7(b).  The fitting range barely satisfies $T < |\Delta|$ near $h_{\perp} \sim 5.3$ Tesla, limiting the accuracy of our estimation of $\Delta$.  To improve the accuracy, we constructed  the scaling plots of $T^{+0.75}/T_{1}$ as a function of $\Delta/T$ in Fig.\ 8.  We first estimated the magnitude of $\Delta$ from Fig.\ 7(a).  Subsequently, for the field range between 5.0 and  6.7 T, we made slight adjustments to the magnitude of $\Delta$ to improve the scaling collapse in Fig.\ 8.  The final results of $\Delta$ thus estimated from Fig.\ 8 are presented in Fig.\ 7(b) using $\blacktriangle$.  We note that this procedure changes the estimated value of  $\Delta$ only by a few K.

Remarkably, we found that $\Delta$ varies linearly with $h_{\perp}$.  This linear behavior is precisely what we expect from the theoretical prediction for the nearest-neighbor quantum Ising chain, $\Delta = 2J(1 - h_{\perp}/h_{\perp}^{c})$ \cite{Sachdev}.  From the intercept of the linear fit with the horizontal axes, we estimate $h_{\perp}^{c} = 5.25 \pm 0.15$ Tesla.  This 1D critical field observed by our NMR measurements agrees very well with the earlier observation of the saturation of the so-called E8 golden ratio \cite{Coldea}.  From the intercept of the linear fit with the vertical axis, we also estimate $J =17.5^{+2.5}_{-1.5}$ K, in excellent agreement with earlier reports based on ESR \cite{ESR} and neutron scattering \cite{Coldea}.  

\subsection{\label{sec:level2}Phase Diagram of the TFIC in CoNb$_2$O$_6$}
We present the color plot of $1/T_{1}$ in Fig.\ 9.  Also shown in Fig.\ 9 is the crossover temperatures, $\Delta$ and $|\Delta|$, based on the linear fit in Fig.\ 7(b).  Our color plot visually captures the crossover from the QC regime to the RC and QD regimes successfully.  We are the first to verify the theoretical $T-h_{\perp}$ phase diagram in Fig.\ 1 for finite temperatures, $T > 0$, using an actual material.

\subsection{\label{sec:level2}Quantum Criticality of the TFIC at Finite Temperatures}
Having established the phase diagram of the TFIC in CoNb$_2$O$_6$, we are ready to test the finite temperature properties of the QC regime located between the RC and QD regimes.  At the 1D critical field $h_{\perp}^{c}$, we applied QC scaling to eq.(3), and obtained 
\begin{equation}
	1/T_{1} = 2.13~ |a_{hf}|^{2} J^{-0.25}T^{-0.75},
\label{3}	
\end{equation}
for the nearest-neighbor TFIC (see eq.\ (A7) below for the details).  We determined the hyperfine form factor $|a_{hf}|^{2}$ based on the $^{93}$Nb NMR frequency shift measurements, and used eq.\ (4) to estimate $1/T_{1} = (4.2 \sim 8.4)\times 10^{3}~T^{-0.75}$ s$^{-1}$ at finite temperatures above the QCP {\it without any adjustable parameters}.  We refer readers to Appendix B for the details of the data analysis.  This parameter-free prediction is in excellent quantitative agreement with our experimental finding, $1/T_{1} \sim 6.2 \times10^{3}~T^{-0.75}$ s$^{-1}$ as shown by a solid line in Fig.\ 5 through the data points observed at 5.2\ T.  Thus the QC scaling theory accounts for the low frequency spin dynamics of the TFIC above $T=0$ at a quantitative level.  

It is equally important to realize that $1/T_{1}$ data exhibits the expected power-law behavior, $1/T_{1} \sim T^{-0.75}$, up to $\sim 7$ K, which corresponds to $T \sim 0.4J$.  Our finding therefore addresses an important and unresolved question that has been facing the strongly correlated electrons community for years: {\it How high in temperature does the effect of the QCP persist?}  For the TFIC, the quantum fluctuations originating from the zero temperature QCP persist up to as high as $T \sim 0.4J$.  Our experimental finding is consistent with the earlier theoretical report that the QC scaling holds up to $T\sim 0.5J$ for the TFIC \cite{Chakravarty2}.    

\section{\label{sec:level1}Summary and conclusions}
Using the quasi one-dimensional Co chains in CoNb$_2$O$_6$, we experimentally tested the quantum criticality of the transverse field Ising chain (TFIC) at finite temperatures above $T=0$ for the first time.   Based on the measurements of the $^{93}$Nb longitudinal relaxation rate $1/T_{1}$, we identified the distinct behaviors of low-frequency spin fluctuations in the Renormalized Classical (RC), Quantum Critical (QC), and Quantum Disordered (QD) scaling regimes of the TFIC, and constructed the $T-h_{\perp}$ phase diagram of the TFIC in Fig.\ 9.  We observed no evidence for a crossover into the 3D regime in the temperature and field range of our concern.  We also reported the transverse field ($h_{\perp}$) dependence of the spin excitation gap parameter $\Delta$ in Fig.\ 7(b); our results exhibit a linear dependence on $h_{\perp}$, in agreement with the theoretical prediction for the nearest-neighbor TFIC.  Our $1/T_{1}$ data observed for the QC regime near $h_{\perp}^{c} \approx 5.25$ T exhibit the expected mild power law divergence, $1/T_{1} \sim T^{-0.75}$ toward the quantum critical point at $T=0$.  Furthermore, the parameter-free prediction based on quantum critical scaling reproduces the magnitude of $1/T_{1}$ within $\sim \pm 36$ \%.  Our results in Fig.\ 5 establish that the quantum critical behavior persists to as high as $T \sim 0.4 J$.  To the best of our knowledge, this is the first example of the quantitative test of the finite temperature effects on quantum criticality for model Hamiltonians with a rigorously solvable ground state.   

We mark the upper bound of the QC scaling regime, $T \sim 0.4 J$, in Fig.\ 9 with a horizontal arrow.  Such a robust quantum criticality observed at finite temperatures above the QCP is in stark contrast with the case of thermally induced {\it classical} phase transitions; the critical region of the latter generally narrows as the phase transition temperature approaches zero, and eventually diminishes at $T=0$ \cite{Lonzarich}.  Many authors have constructed analogous color plots for different parameters (such as electrical resistivity, as an example) for a variety of strongly correlated electron systems, including copper-oxide and iron-pnictide high $T_c$ superconductors and heavy Fermion systems \cite{Sachdev3, Si}.  The aim of these authors was to build a circumstantial case that quantum fluctuations persist at finite temperatures far above the QCP.  The overall similarity between our Fig.\ 9 and the case of high $T_c$ cuprates and other exotic superconductors gives us hope that quantum fluctuations may indeed account for the mechanism of exotic superconductivity.\\

Note Added: After the initial submission of this work, a theoretical prediction was made for the temperature dependence of $1/T_{1}$ under the presence of an internal longitudinal magnetic field in the three-dimensionally ordered state \cite{Wu}.  The three-dimensional effects \cite{Balents, Wu}, however, are beyond the scope of the present work.  

\begin{acknowledgments}
T.I. and S.S. thank helpful communications with A. P. Young, Y. Itoh, B. Gaulin, M. P. Gelfand, S.-S. Lee, T. Sakai and H. Nojiri.  The work at McMaster was supported by NSERC and CIFAR.  S.S. acknowledges the financial support from NSF DMR-1103860.\\
\end{acknowledgments}
\appendix

\section{\label{sec:level1}Theoretical derivations of $1/T_{1}$ in the quantum Ising chain}

Here we will summarize the derivations of the theoretical expressions of $1/T_{1}$ in the TFIC.  Our notation will be the same as in \cite{Sachdev}.  Some results will be specific to the nearest-neighbor Ising model, but most are more generally applicable to the vicinity of the quantum critical point of a generic one-dimensional Ising chain.    In general, the NMR relaxation rate is defined by
\begin{subequations}
\begin{align}
\frac{1}{T_{1}} = \lim_{\omega \to 0} \frac{2T}{\omega} \int_{}^{} \frac{dk}{2 \pi}~ |a_{hf}|^{2}~  \text{Im} \chi(k, \omega)\\
=  \int_{}^{} \frac{dk}{2 \pi}~ |a_{hf}|^{2}~ \text{S}(k, \omega=0)\\
= \int_{-\infty}^{+\infty} dt~ |a_{hf}|^{2}~ \text{C}(x=0, t),
\end{align}
\end{subequations}
where $a_{hf}$ represents the hyperfine coupling between the nuclear spin and the Pauli matrices $\sigma$, as defined by the hyperfine Hamiltonian $\hat{H}_{hf} = \hat{I} \cdot a_{hf}\cdot \hat{\sigma}$.  We define the correlation function for Pauli matrices, and $\hbar = k_{B} = 1$ unless noted otherwise.
  
\subsubsection{Renormalized Classical Regime}
This region is characterized by an energy gap $\Delta \sim (g_{c}-g)$ and a $T=0$ ordered moment $N_{o} \sim (g_{c}-g)^{1/8}$.    The  $N_{o}$ represents the ordered moment of an Ising chain at $T=0$, and should not be confused with the 3D ordered moment induced by inter-chain couplings.  By expressing our results in terms of $\Delta$ and $N_{o}$, they are generally valid {\it beyond} the nearest-neighbor model.  For the specific case of the nearest-neighbor model, we have $\Delta = 2J(1-g)$ and $N_{o} = (1-g^{2})^{1/8}$.  The result for $C(x, t)$ may be found below (4.81) in Ref. \cite{Sachdev}, and this leads to 
\begin{equation}
\frac{1}{T_{1}} = |a_{hf}|^{2} \frac{\pi N_{o}^{2}}{T} e^{+\Delta/T}.
\end{equation}
Notice that $1/T_{1}$ is expected to diverge exponentially, even though there is an energy gap $\Delta$ in the excitation spectrum of the domain-wall quasi-particles.  This is because NMR is a low energy probe, and $1/T_{1}$ in the RC regime is dominated by the low frequency spin fluctuations associated with the quasi-elastic mode of the 1D Ising chain induced by ferromagnetic short range order.  

Our scaling analysis in Fig.\ 8(a) suggests that the observed divergent behavior of $1/T_{1}$ is somewhat weaker than $\frac{1}{T_{1}} \sim \frac{1}{T} e^{+\Delta/T}$, perhaps because our experimental range of $T$ and $h_{\perp}$ is not deep inside the RC regime, or possibly due to the influence of additional terms in the Hamiltonian neglected in the theoretical calculations.  Accordingly, we fit the $1/T_{1}$ data in the RC regime with the simple exponential form, $1/T_{1} \propto e^{+\Delta/T}$, in Fig.\ 7(a), ignoring the temperature dependent pre-factor $\sim 1/T$.  

\subsubsection{Quatum Critical Regime}
Here, we have in imaginary time, $\tau$, from (4.106) in Ref. \cite{Sachdev} that 
\begin{equation}
C(x=0, \tau) =  Z T^{1/4} \frac{G_{I}(0)}{[2 \sin(\pi T \tau)]^{1/4}},
\end{equation}
where $G_{I}(0) = 0.858714569$, and
\begin{equation}
Z =  \lim_{\Delta \to 0} \frac{N_{o}^{2}}{\Delta^{1/4}};
\end{equation}
the value of $Z$ is a general result upon approaching from the ordered side, valid beyond the nearest-neighbor model.

From eq.\ (A3), we have the local susceptibility in imaginary time
\begin{equation}
\chi(x=0, \omega_{n}) = \int_{0}^{1/T} d\tau ~e^{i \omega_{n} \tau} C(x=0, \tau).
\end{equation}
We evaluate the Fourier transform using (3.12), (3.22), and (3.24) of Ref. \cite{Senthil}, and obtain

\begin{eqnarray}
\text {Im}~ \chi(x=0, \omega_{n}) = \frac{ZG_{I}(0)}{T^{3/4}2^{1/4}\sqrt{\pi} \Gamma(1/8) \Gamma(5/8)} \nonumber \\
\times \sinh (\frac{\omega}{2T}) |\Gamma (\frac{1}{8}-\frac{i \omega}{2\pi T})|^{2}.
\end{eqnarray}

This gives us
\begin{equation}
\frac{1}{T_{1}} = |a_{hf}|^{2} \frac{Z}{T^{3/4}} \frac{G_{I}(0) \Gamma(1/8)}{2^{1/4} \sqrt{\pi} \Gamma(5/8)} = 2.13~|a_{hf}|^{2}~ \frac{Z}{T^{3/4}}.
\end{equation} \\
In the case of the nearest-neighbor Ising model, $\Delta = 2J (1- g)$ and $N_{o} = (1-g^{2})^{1/8}$.  Accordingly,  we obtain $Z=J^{-1/4}$ from eq.\ (A4), and hence eq.\ (A7) leads to eq.\ (4) in the main text.  

 \subsubsection{Quatum Disordered Regime} 
Here we can expect that $1/T_{1}$ diminishes exponentially in the quantum disordered regime due to the excitation gap, $|\Delta|$, and so
\begin{equation}
\frac{1}{T_{1}} \propto e^{-|\Delta|/T},
\end{equation} where now $\Delta < 0$.  However there is no explicit computation in the TFIC establishing this,
and the pre-factor is unknown. Accordingly, we fit the $1/T_{1}$ data in Fig.\ 7(a) to the simple activation form.

\section{\label{sec:level1}Analysis of $1/T_{1}$ in the QC Regime}
In the previous section, we defined the hyperfine coupling with Pauli matrices as $a_{hf}$ to maintain consistency of the notation for dynamical spin susceptibility defined in \cite{Sachdev}.  To use the standard notations of NMR data analysis, here we introduce the hyperfine coupling $A_{hf}$ between the nuclear spin $I$ and electron spin $S$ through the hyperfine Hamiltonian $\hat{H}_{hf} = \hat{I} \cdot A_{hf}\cdot \hat{S}$.  That is, $a_{hf} = SA_{hf}$.  Earlier ESR measurements determined the anisotropic g-tensor of the Co$^{2+}$ ions in CoNb$_2$O$_6$ as $g^{(a)} = 4.3$ and $g^{(c)} = 6.1$ by taking the Co pseudo spin as $S=\frac{1}{2}$ \cite{ESR}.

Recalling that $1/T_{1}$ measured with an external magnetic field applied along the crystal b-axis probes the fluctuating hyperfine fields along the a- and c-axes, we may rewrite eq.\ (A7) as     
\begin{equation}
\frac{1}{T_{1}} = 2.13~S^{2}~ \frac{|A_{hf}^{(a)}/\hbar|^{2}+|A_{hf}^{(c)}/\hbar|^{2}}{2}~ \frac{\hbar}{(k_{B}J)^{1/4}(k_{B}T)^{3/4}},
\end{equation} 
where we show $\hbar$ and $k_{B}$ explicitly.  

Next, we estimate the uniform ${\bf k} = {\bf 0}$ component of the hyperfine coupling from the NMR frequency shift $K$ \cite{Jaccarino}  
\begin{equation}
K^{(\alpha)} = \frac{A_{hf}^{(\alpha)}({\bf k} = {\bf 0})}{g^{(\alpha)} \mu_{B}} \chi^{(\alpha)} + K_{chem}^{(\alpha)},
\end{equation} 
where $\alpha =$ a, b, and c, and $K_{chem}^{(\alpha)}$ is the small temperature independent chemical shift.  Accordingly,
\begin{equation}
\frac{A_{hf}^{(\alpha)}({\bf k} = {\bf 0})}{\hbar} = \gamma_{n} N_{A} g^{(\alpha)} \mu_{B}  ~ \frac{dK^{(\alpha)}}{d\chi^{(\alpha)}},
\end{equation} 
where the $^{93}$Nb nuclear gyromagnetic ratio is $\gamma_{n}/2\pi =10.407$~MHz/Tesla, and $N_{A}$ is Avogadro's number.  

To determine the only unknown parameter $\frac{dK^{(\alpha)}}{d\chi^{(\alpha)}}$ in the right hand side of eq.\ (B3), we plot $K^{(\alpha)}$ in Fig.\ 10 as a function of the molar magnetic susceptibility $\chi^{(\alpha)}$ measured along the corresponding orientations (see Fig.\ 2(d)), choosing $T$ as the implicit parameter.  From the  linear fit of the $K$ vs. $\chi$ plot, we estimate the slope as $\frac{dK^{(\alpha)}}{d\chi^{(\alpha)}} = 0.386$, 0.221, and 0.311 for $\alpha = a$, b, and c, respectively.   Therefore we arrive at  $A_{hf}^{(a)}({\bf k} = {\bf 0})/\hbar = 6.0 \times 10^{7}$ (s$^{-1}$) and $A_{hf}^{(c)}({\bf k} = {\bf 0})/\hbar = 7.0 \times 10^{7}$ (s$^{-1}$).

Next, we need to relate these results with the fluctuating hyperfine fields $|A_{hf}^{(\alpha)}/\hbar|^{2}$ in eq.\ (B1).  The upper bound of the latter may be easily estimated as, 
\begin{equation}
|A_{hf}^{(\alpha)}/\hbar|^{2} = |A_{hf}^{(\alpha)}({\bf k} = {\bf 0})/\hbar|^{2},
\end{equation} 
where we assumed that all Co chains fluctuate coherently with ferromagnetic inter-chain correlations.  Inserting eq.\ (B4) into eq.\ (B1), we obtain $1/T_{1} = 8.4 \times 10^{3} ~ T^{-0.75} $ (s$^{-1}$).  This theoretical upperbound overestimates the experimental results observed for $\sim 5.2$\ T by $\sim 36$ \%.      

In reality, the inter-chain couplings are smaller than $J$ by an order of magnitude, and frustrated.  Since we are concerned with the temperature range $T > 0.1 J$, it is safe to assume that the fluctuating transferred hyperfine fields from two nearby Co-O-Co chains are uncorrelated.  Assuming that the magnitude of these couplings are comparable ($\sim A_{hf}^{(\alpha)}({\bf k} = {\bf 0})/2\hbar$), and that their fluctuations are additive, we arrive at
\begin{equation}
|A_{hf}^{(\alpha)}/\hbar|^{2} \sim 2 \times |A_{hf}^{(\alpha)}({\bf k} = {\bf 0})/2\hbar|^{2}.
\end{equation} 
By inserting eq.\ (B5) into eq.\ (B1), we estimate  $\frac{1}{T_{1}} = 4.2 \times 10^{3} ~T^{-0.75}$ (s$^{-1}$).  This underestimates the experimental observation by  $\sim 33$ \%.  


\pagebreak[1]

\begin{figure} \centering
\includegraphics[width=3.2in]{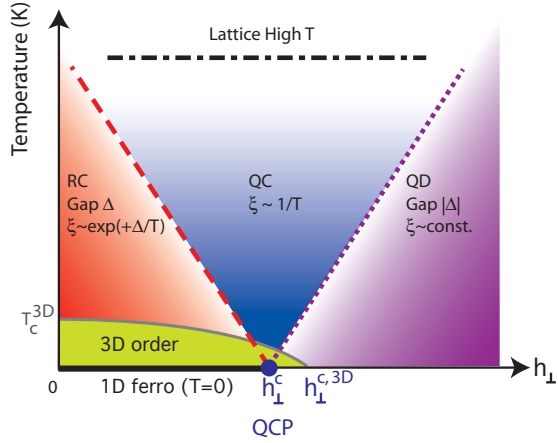}
\caption{\label{Fig1:epsart} A generic $T - h_{\perp}$ phase diagram of the TFIC encompasses three scaling regimes with distinct behaviors of the spin-spin correlation length $\xi$: RC (Renormalized Classical, $g<1$ hence $h_{\perp} < h_{\perp}^{c}$, and $\xi \sim \exp(+\Delta/T)$), QC (Quantum Critical, $\xi \sim 1/T$), and QD (Quantum Disordered, $g>1$ hence $h_{\perp} > h_{\perp}^{c}$, and $\xi \sim$ constant) \cite{Sachdev}. The dashed and dotted lines represent the crossover temperature from the QC to RC regime at $T\sim \Delta$ and from the QC to QD regime at $T\sim |\Delta|$, respectively.  An isolated 1D Ising chain would exhibit ferromagnetic long range order only at $T=0$ below $h_{\perp}^{c}$, but the 3D inter-chain couplings lead to a 3D order at $T>0$ up to $h_{\perp}^{c, 3D}$ ($> h_{\perp}^{c}$).  The filled circle at $T=0$ and the 1D (one dimensional) critical field $h_{\perp}^{c}$ represents the quantum critical point (QCP) of the individual Ising chain.}     
\end{figure}

\begin{figure}\centering
\includegraphics[width=3.2in]{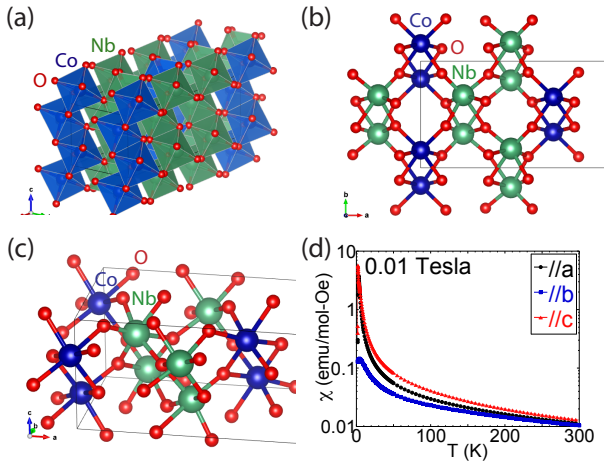}
\caption{\label{Fig2:epsart} (a) The crystal structure of CoNb$_2$O$_6$.  (b) Both magnetic CoO$_{6}$ and non-magnetic NbO$_{6}$ octahedra form a chain along the c-axis, as seen from the c-axis direction.  The Nb-O-Nb chain is inside an isosceles triangle formed by three Co-O-Co chains.  The transverse field $h_{\perp}$ is applied along the b-axis.  (c) Each Nb site is bonded with two Co-O-Co chains across O sites.  (d) Bulk magnetic susceptibility $\chi$ data measured with SQUID in an external magnetic field of 0.01 T.}     
\end{figure}

\begin{figure} \centering
\includegraphics[width=3.2in]{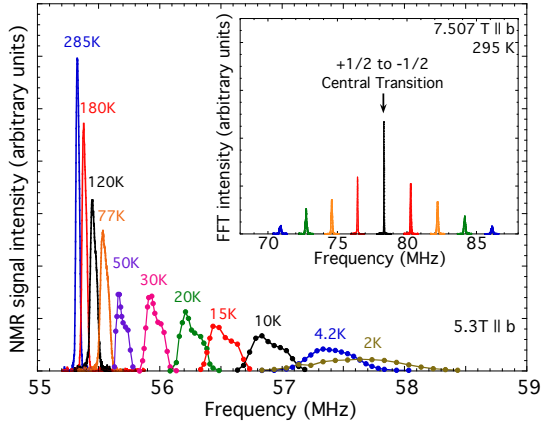} 
\caption{\label{Fig3:epsart}  The temperature dependence of the $^{93}$Nb NMR lineshape observed for the central transition between the $I_{z} = +\frac{1}{2}$ and $-\frac{1}{2}$ energy levels in $h_{\perp} = 5.3$ Tesla applied along the b-axis.  We obtained the lineshapes using the FFT of the spin echo signal above 77\ K.  For the broader lineshapes below 77\ K, we measured the integral of the spin echo as a function of the frequency.  Inset: the $^{93}$Nb NMR lineshape at 295\ K observed at 7.507\ T usings the FFT of spin echo signals.  The largest peak in the middle is the central transition, and four additional pairs of weaker peaks arise from $I_{z} = m$ to $m+1$ transitions ($m=-9/2$, -7/2, -5/2, -3/2, +1/2, +3/2, +5/2, and +7/2).}      
\end{figure} 

\begin{figure} \centering
\includegraphics[width=3.2in]{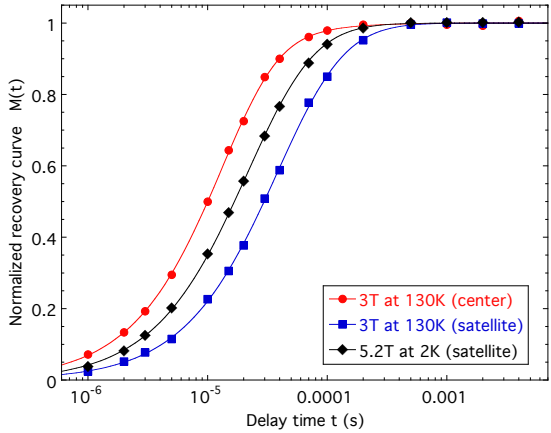} 
\caption{\label{Fig4:epsart} Examples of the recovery of the spin echo intensity, $M(t)$, observed for the central and a fourth satellite transition at 130 K in $h_{\perp} = 3$ Tesla.  For comparison, we normalized the recovery curves by plotting $1-[M(\infty)-M(t)]/A$ as a function of $t$.  The solid lines represent the best fit with $1/T_{1}=1.99 \times 10^{3}$ s$^{-1}$ for the central transition and $1/T_{1}=1.96 \times 10^{3}$ s$^{-1}$ for the fourth satellite transition, as described in the text.  Also plotted is the recovery curve observed for the fourth satellite peak at 2 K in $h_{\perp} = 5.2$ Tesla.}     
\end{figure}  

\begin{figure}\centering
\includegraphics[width=3.2in]{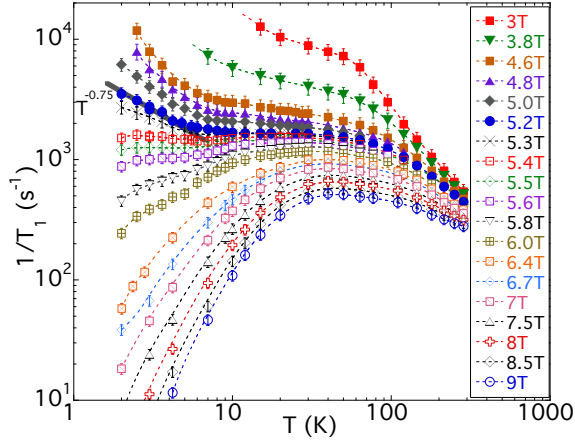} 
\caption{\label{Fig5:epsart} The temperature dependence of $1/T_{1}$ in $h_{\perp}$ applied along the b-axis. All dashed lines interconnecting the data points are guides for the eye.  The black solid line through the 5.2 T data points  represents a power-law fit, $1/T_{1} \sim 6.2 \times10^{3}~T^{-0.75}$ s$^{-1}$.}     
\end{figure}

\begin{figure}\centering
\includegraphics[width=3.2in]{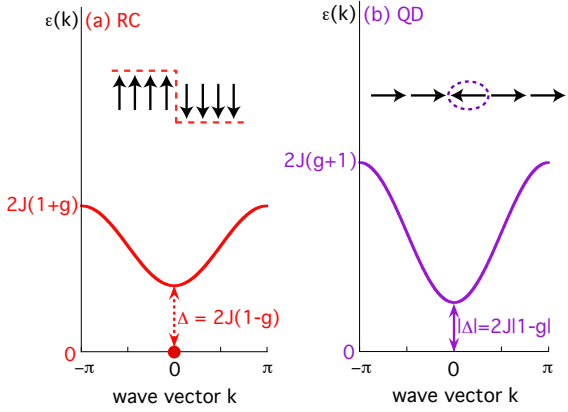}
\caption{\label{Fig6:epsart}  (a) The spin excitation spectrum in the RC regime has two components, the quasi-elastic peak at the origin (represented by a filled dot) and the propagating domain walls, as schematically shown in the inset.  The dispersion of the latter (solid curve) is $\epsilon (k) = J [2 - 2g \cos(k) + O(g^{2})]$, with an excitation gap $\Delta = 2J(1-g)$ \cite{Sachdev}.  The quasi-elastic peak becomes a Bragg peak when $\xi$ diverges toward the 1D ferromagnetic long range order at $T=0$.  Since NMR is a low energy probe, our $1/T_{1}$ data measured below $h_{\perp}^{c}$ probe the quasi-elastic mode.  (b) The spin excitation spectrum in the QD regime, $\epsilon (k) = J g [2 - (2/g) \cos(k) + O(1/g^{2})]$ with a gap $| \Delta | = 2 | 1-g |$ \cite{Sachdev}, arises from the propagation of flipped spins (inset).  Unlike the RC regime, there is no quasi-elastic peak.}     
\end{figure}

\begin{figure}\centering
\includegraphics[width=3.2in]{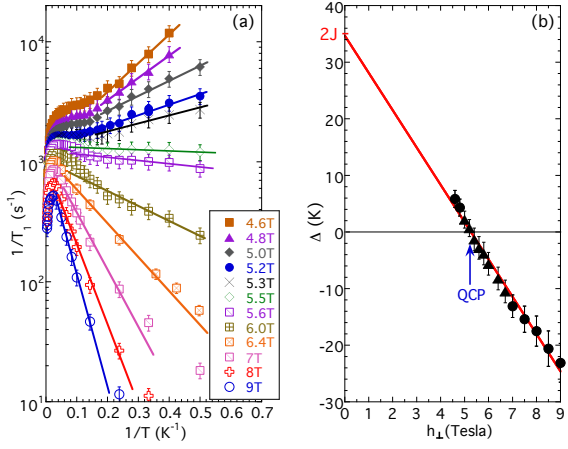} 
\caption{\label{Fig7:epsart} Estimation of the gap $\Delta$.  (a) The exponential fit $1/T_{1} \sim \exp(\Delta/T)$ for representative values of $h_{\perp}$.  (b) $\bullet$ represents $\Delta$ as determined from (a), while $\blacktriangle$ is based on the scaling analysis.  Also shown is a linear fit, $\Delta = 2J(1- h_{\perp}/h_{\perp}^{c})$.  From the fit, we estimate $h_{\perp}^{c} = 5.25 \pm 0.15$ Tesla and $J = 17.5 ^{+2.5}_{-1.5}$ K.}     
\end{figure}

\begin{figure} \centering
\includegraphics[width=3.2in]{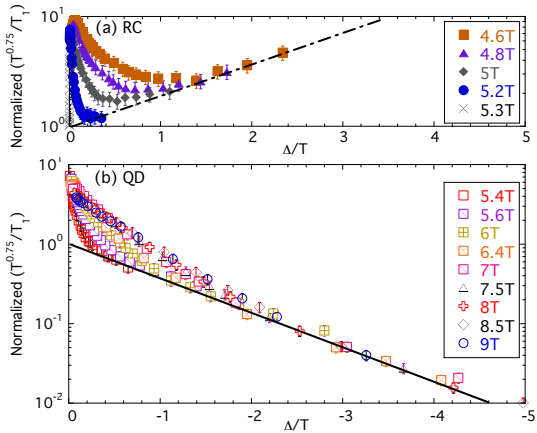} 
\caption{\label{Fig8:epsart} The scaling plots of $T^{+0.75}/T_{1}$ as a function of $\Delta/T$ in (a) the RC regime, and (b) the QD regime.  For clarity, we normalized the overall magnitude of $T^{+0.75}/T_{1}$ as unity for the QC regime.  The dashed-dotted line in (a) is a guide-for-eyes, while the solid line in (b) represents $1/T_{1} \propto e^{-| \Delta |/T}$.}     
\end{figure} 

\begin{figure}\centering
\includegraphics[width=3.2in]{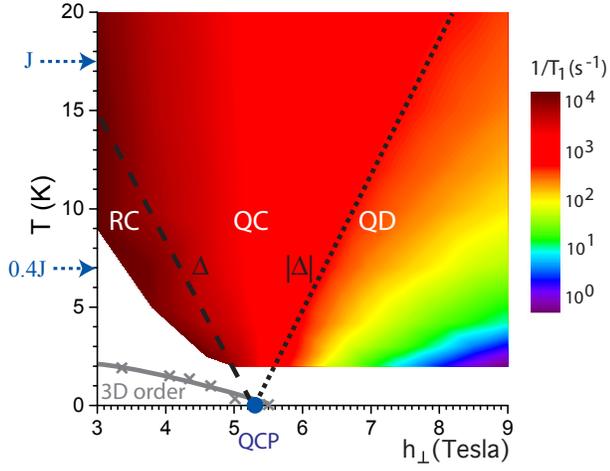} 
\caption{\label{Fig9:epsart} A color plot of  $1/T_{1}$.  The dashed (dotted) line represents the expected crossover temperature $\Delta$ ($|\Delta|$) from the QC to RC (QD) regime, based on the linear $h_{\perp}$ dependence of $\Delta$ estimated in Fig.\ 7(b).  Also shown (grey x) is the 3D ordering temperature $T_{c}^{3D}$ \cite{Wheeler}.}     
\end{figure}

\begin{figure} \centering
\includegraphics[width=3.2in]{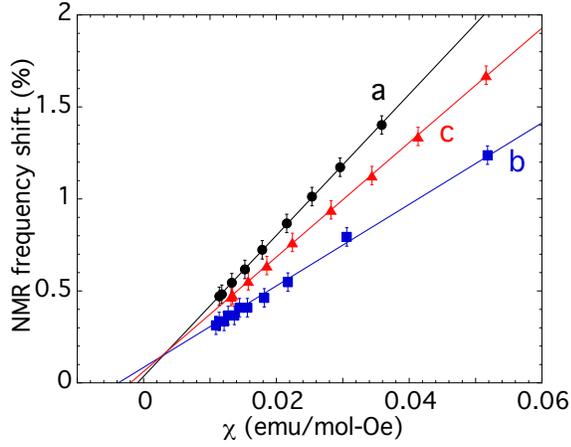} 
\caption{\label{Fig10t:epsart} The NMR frequency shift $K^{(\alpha)}$ vs. the bulk magnetic susceptibility $\chi^{(\alpha)}$, with $T$ as the implicit parameter ($\alpha = a$, b, or c).  The straight lines are the best linear fits.}     
\end{figure}

\end{document}